\documentstyle[amsfonts,twocolumn,prl,aps,epsfig]{revtex}
\begin{document}
\twocolumn[\hsize\textwidth\columnwidth\hsize\csname
@twocolumnfalse\endcsname
\draft
\title{Entanglement degree for Jaynes-Cummings model}
\author{S. Furuichi}
\address{Science University of Tokyo in Yamaguchi, \\ 1-1-1 Daigakudori Onoda city Yamaguchi, 756-0884, Japan}
\author{M. Ohya}
\address{Department of Information Sciences, Science University of Tokyo, \\
2641 Yamazaki Noda city Chiba, 278-8510, Japan}
\date{\today}
\vspace{1cm}
\maketitle
\vspace{0.5mm}
\begin{abstract}
\quad Recently, it has been known that a quantum entangled state plays an important role in the field of quantum information theory such as quantum teleportation and quantum computation. The research on quantifying entangled states has been done by several measures. In this letter, we will adopt the method using quantum mutual entropy to measure the degree of entanglement on the Jaynes-Cummings model.
\end{abstract}
\vspace{0.5mm}
\pacs{PACS numbers: 89.70.+c, 03.65.Bz, 42.50.Dv}
\vspace{0.5mm}]

The degree of entanglement for mixed states has been studied by some entropic
measures such as entanglemnt of formation\cite{Ben}, quantum relative entropy
\cite{Ved} and the degree of entanglement which is defined by entropy minus  
mutual entropy\cite{BO}. In this letter, we will
apply the degree of entanglement due to quantum mutual entropy\cite{O1}, we
call it DEM in the sequel, to the entangled state in the Jaynes-Cummings model
\cite{JC}. Vedral {\it et al.} defined the degree of entangled state $\sigma
$ as a minimum distance between all disentangled states $\rho \in {\cal D}$
such that $E\left( \sigma \right) \equiv {\min_{_{\rho \in {\cal D}}}}
D\left( \sigma ||\rho \right) $ where $D$ is any measure of distance between
the two states $\sigma $ and $\rho $\cite{Ved}. For an example, one can choose 
quantum relative entropy as $D$. Then,
\begin{equation}
E\left( \sigma \right) ={\min_{\rho \in {\cal D}}}S\left( \sigma |\rho
\right) ,  \label{QRE}
\end{equation}
where $S\left( \sigma |\rho \right) \equiv tr\sigma \left( \log \sigma -\log
\rho \right) $ is quantum relative entropy\cite{U,OP}. Since this measure has
to take a minimum over all disentangled states, it is difficult to calculate
analytically for the Jaynes-Cummings model so that we use the DEM defined below. Moreover, there has been no fixed-definition of entanglement measure, though some measures have been defined other than the above measure defined in (\ref{QRE}).
So we can use the convenient measure which ever we want case-by-case.

Let $\sigma $ be a state in ${\frak S}_{1}\otimes {\frak S}_{2}$ and $\rho
_{k}$ are the marginal states in ${\frak S}_{k}$ ({\it i.e.} $tr_{j}\sigma
=\rho _{k}\left( k\neq j\right) $ ). Then our degree of entanglement due to quantum mutual entropy (DEM) is defined by:
$$I_{\sigma }\left( \rho _{1},\rho _{2}\right) \equiv tr\sigma \left( \log
\sigma -\log \rho _{1}\otimes \rho _{2}\right).$$
Note that the tensor product state $\rho _{1}\otimes \rho _{2}$ is one of the
disentangled states. This quantity is applied to classify entanglements in
\cite{BO}. If $\sigma \in {\frak S}_{1}\otimes {\frak S}_{2}$ is an entangled
pure state, then its von Neumann entropy is equal to $0$ ({\it i.e.}, $S\left( \sigma \right) =0$). Moreover,
from the following inequality\cite{AL}:
$$|S\left( \rho _{1}\right) -S\left( \rho _{2}\right) |\leq S\left( \sigma
\right) \leq S\left( \rho _{1}\right) +S\left( \rho _{2}\right)$$
we have $S\left( \rho _{1}\right) =S\left( \rho _{2}\right) $. Thus we have
\begin{eqnarray}
I_{\sigma }\left( \rho _{1},\rho _{2}\right) &=&tr\sigma \left( \log \sigma
-\log \rho _{1}\otimes \rho _{2}\right)  \nonumber \\
&=&S\left( \rho _{1}\right) +S\left( \rho _{2}\right) -S\left( \sigma \right)
\label{DEM} \\
&=&2S\left( \rho _{1}\right)  \nonumber
\end{eqnarray}
Therefore, for entangled pure states, the DEM becomes twice of von Neumann
reduced entropy. That is, if we want to know the degree of the entangled
pure states, it is sufficient to use von Neumann reduced entropy. However,
for entangled mixed states which appear in many cases, we have to use the DEM.
From (\ref{DEM}), we also find that the entanglement degree in a pure state is bigger than that in a mixed state. 

The Jaynes-Cummings model has been studied by many researchers from various
points of view\cite{YE,SK} because this model has a very simple form and can
be exactly solvable. One of the most interesting features on this model is the
entanglement developed between the atom and the field during the
interaction. There have been several approaches to analyze the time
evolution in this model, for instance, von Neumann entropy and atomic
inversion. In this letter, we will apply the DEM to analyze the entanglement
of the Jaynes-Cummings model.

The resonant Jaynes-Cummings model Hamiltonian can be expressed by rotating-wave approximation
in the following form
\begin{eqnarray}
H &=&H_{A}+H_{F}+H_{I},  \nonumber \\
H_{A} &=&{\frac{1}{2}}\hbar \omega _{0}\sigma _{z},\,H_{F}=\hbar \omega
_{0}a^{*}a,  \nonumber \\
H_{I} &=&\hbar g(a\otimes \sigma ^{+}+a^{*}\otimes \sigma ^{-}),
\nonumber
\end{eqnarray}
where $g$ is a coupling constant, $\sigma ^{\pm }$ are the pseudo-spin
matrices of two-level atom, $\sigma _{z}$ is the $z$-component Pauli spin 
matrix, $a$ (resp. $a^{*}$) is the annihilation (resp. creation) operator of a photon.
We now suppose that the initial state of the atom is a superposition of the grounded state and the excited state:
$$\rho =\lambda _{0}E_{0}+\lambda _{1}E_{1}\in {\frak S}_{A}  $$
where $E_{0}=\left| 1\right\rangle \left\langle 1\right| $, $E_{1}=\left|
2\right\rangle \left\langle 2\right| $, $\lambda _{0}+\lambda _{1}=1$. 
Let the field initially be in a coherent state:
$$\omega =\left| \theta \right\rangle \left\langle \theta \right| \in {\frak S}
_{F},\,\left| \theta \right\rangle =\exp \left( -\frac{1}{2}|\theta
|^{2}\right) \sum_{l}\frac{\theta ^{l}}{\sqrt{l!}}|l\rangle $$
The continuous map ${\cal E}_{t}^{*}$ describing the time evolution between
the atom and the field for the Jaynes-Cummings model is defined by the unitary operator generated by $H$ such that
$${\cal E}_{t}^{*}:{\frak S}_{A}\longrightarrow {\frak S}_{A}\otimes {\frak S}
_{F},$$
\begin{equation}
{\cal E}_{t}^{*}\rho =U_{t}\left( \rho \otimes \omega \right) U_{t}^{*},
\label{ent_state}
\end{equation}
$$U_{t}\equiv \exp \left( -it\frac{H}{\hbar }\right).$$
This unitary operator $U_{t}$ is written as
\begin{equation}
U_{t}=\exp \left( {-itH/\hbar }\right) =\sum\limits_{n=0}^{\infty }{%
\sum\limits_{j=0}^{1}E_{n,j}\left| {\Phi _{j}^{\left( n\right) }}%
\right\rangle \left\langle {\Phi _{j}^{\left( n\right) }}\right| },
\label{unitary}
\end{equation}
where $E_{n,j}=\exp \left[ -it\left\{ \omega _{0}\left( n+\frac{1}{2}\right)
+\left( -1\right) ^{j}\Omega _{n}\right\} \right] $ are the eigenvalues with $
\Omega _{n}=g\sqrt{n+1},$ called Rabi frequency and ${\Phi _{j}^{\left(
n\right) }\ }$ are the eigenvectors associated to $E_{n,j}$.

The transition probability which the atom is initially prepared in the excited
state and stays at the excited state after the time $t$ is given by

\begin{eqnarray}
c\left( t \right)&=&\left| {\left\langle {n\otimes 2} \right|U_t\left| {%
n\otimes 2} \right\rangle } \right|^2  \nonumber \\
&=&\exp\left(-\vert\theta\vert^2\right)\sum\limits_n{\ \frac{%
\vert\theta\vert^{2n}}{n!} \cos ^2 {\Omega _nt} }. \nonumber
\end{eqnarray}

\begin{figure}[htbp]
\begin{center}
\includegraphics[width=8cm]{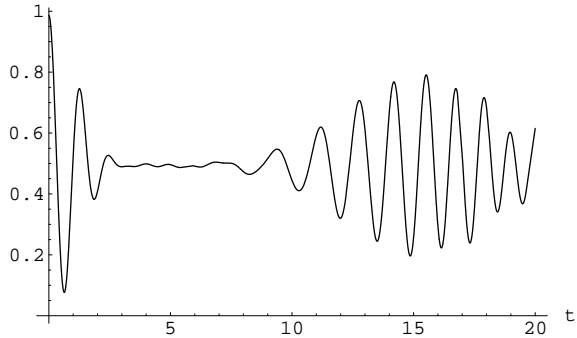}
\caption{Transition probability as a function of time $t$.}
\end{center}
\end{figure}

FIG.1 shows the transition probability for a coupling constant $g=1$ and a
mean photon number $|\theta |^{2}=5$. This model yields a {\it dephase}
around the time $t_{c}\approx 1/g$ and shows the damped oscillation with the 
gaussian envelope. 
This dampping is caused by the difference and interference of the Rabi frequency $\Omega _{n}$. 
This damping phenomenon is often called {\it Cummings collapses}\cite{SK,Cu}. 
FIG.1 shows that the atom in this model is the most uncertain state at the time when the transition probability is equal to $0.5$, namely $t \approx 3 \sim 7$. So, it can be seen that the system in this model at that time becomes the most correlated. 
Later the system shows {\it revival} around $t_{r}\approx 2\pi |\theta |/g$. 
The reason of {\it revival} is considered as a {\it rephase}, and its revival periodically appears at each $T_{k}=kt_{r}$ $\left( k=1,2,3,\cdots \right) $.
It is also known\cite{SK,Ba} that the system in this model returns most closely to a pure state of the atom around $t_r /2$, during the collapse interval.
For details on this model, the readers may refer to the excellent reviews \cite{YE,SK}.

From (\ref{ent_state}) and (\ref{unitary}), the entangled state between the
atom and the field follows as
\begin{eqnarray}
{\cal E}_{t}^{*}\rho  &=&e_{1}\left( t\right) \left| {2\otimes n}%
\right\rangle \left\langle {2\otimes n}\right|   \nonumber \\
&+&e_{2}\left( t\right) \left| {2\otimes n}\right\rangle \left\langle {%
1\otimes n+1}\right|   \nonumber \\
&+&e_{3}\left( t\right) \left| {1\otimes n+1}\right\rangle \left\langle {%
2\otimes n}\right|   \nonumber \\
&+&e_{4}\left( t\right) \left| {1\otimes n+1}\right\rangle \left\langle {%
1\otimes n+1}\right| .  \label{entangle}
\end{eqnarray}
where
\begin{eqnarray}
e_{1}\left( t\right)  &=&\lambda _{0}s{\left( t\right) }+\lambda _{1}c{%
\left( t\right) },  \nonumber \\
e_{2}\left( t\right)  &=&{\frac{i}{2}}\exp \left( -|\theta |^{2}\right)
\left( \lambda _{1}-\lambda _{0}\right) \sum\limits_{n}{\frac{|\theta |^{2n}%
}{n!}\sin 2\Omega _{n}t},  \nonumber \\
e_{3}\left( t\right)  &=&{\frac{{-i}}{2}}\exp \left( -|\theta |^{2}\right)
\left( \lambda _{1}-\lambda _{0}\right) \sum\limits_{n}{\ \frac{|\theta
|^{2n}}{n!}\sin 2\Omega _{n}t},  \nonumber \\
e_{4}\left( t\right)  &=&\lambda _{0}c{\left( t\right) }+\lambda _{1}s{%
\left( t\right) },  \nonumber \\
s{\left( t\right) } &=&\exp \left( -|\theta |^{2}\right) \sum_{n}\frac{%
|\theta |^{2n}}{n!}\sin ^{2}\Omega _{n}t.  \nonumber
\end{eqnarray}
Taking the partial trace of (\ref{entangle}), we have the reduced density
operators $\rho _{t}^{A}$ and $\rho _{t}^{F}$ for the atom and the field,
respectively:
\begin{eqnarray}
\rho _{t}^{A} &=&tr_{F}{\cal E}_{t}^{*}\rho =e_{1}\left( t\right) \left|
2\right\rangle \left\langle 2\right| +e_{4}\left( t\right) \left|
1\right\rangle \left\langle 1\right|   \nonumber \\
\rho _{t}^{F} &=&tr_{A}{\cal E}_{t}^{*}\rho =e_{1}\left( t\right) \left|
n\right\rangle \left\langle n\right| +e_{4}\left( t\right) \left| {n+1}%
\right\rangle \left\langle {n+1}\right| .  \nonumber
\end{eqnarray}
Thus, the degree of the entangled state ${\cal E}_{t}^{*}\rho $ for
Jaynes-Cummings model can be computed as
\begin{eqnarray}
I_{{\cal E}_{t}^{*}\rho }\left( \rho _{t}^{A},\rho _{t}^{F}\right)  &=&tr
{\cal E}_{t}^{*}\rho \left( \log {\cal E}_{t}^{*}\rho -\log \rho
_{t}^{A}\otimes \rho _{t}^{F}\right)   \nonumber \\
&=&S\left( \rho _{t}^{A}\right) +S\left( \rho _{t}^{F}\right) -S\left( {\cal %
E}_{t}^{*}\rho \right)   \nonumber \\
&=&-e_{1}\left( t\right) \log e_{1}\left( t\right) -e_{4}\left( t\right)
\log e_{4}\left( t\right)   \nonumber \\
&+&e_{2}\left( t\right) \log e_{2}\left( t\right) +e_{3}\left( t\right) \log
e_{3}\left( t\right). \nonumber
\end{eqnarray}

FIG.2 shows the time development of the DEM for the Jaynes-Cummings model with the mean photon number $|\theta |^{2}=5$, the coupling constant $g=1$ and the parameters $\left( \lambda _{0},\lambda_{1}\right) =\left( 0.7,0.3\right)$.

\begin{figure}[htbp]
\begin{center}
\includegraphics[width=8cm]{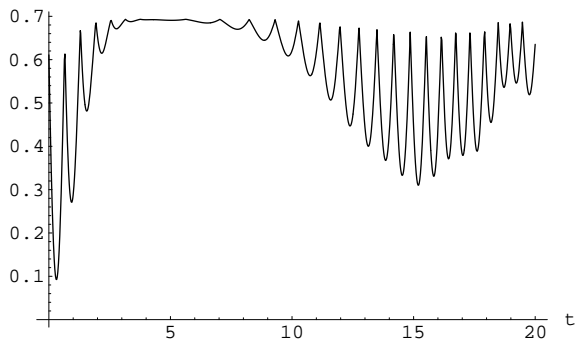}
\caption{The DEM as a function of time $t$.}
\end{center}
\end{figure}

From FIG.2, we find the time-development of the DEM for the Jaynes-Cummings model periodically shows the Rabi oscillation, which is an important feature in this model. Moreover, the entanglement degree takes the highest value around the time $t \approx 3\sim 7$ which corresponds to the collapse interval indicated in FIG.1. 
Therefore we conclude that the system in this model has the strongest entanglement when the system is the most correlated. 
This result also coincides with the fact that the entanglement becomes bigger as the entangled state is closer to a pure state, which we also mentioned above in this letter.

\begin{figure}[htbp]
\begin{center}
\includegraphics[width=8cm]{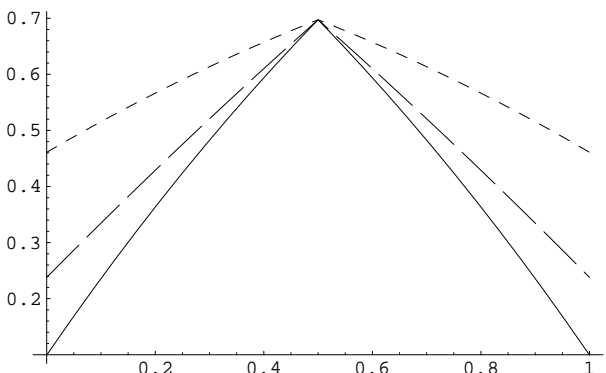}
\caption{The DEM as functions of parameter $\lambda_0$ for $T_1$(solid curve), $T_2$(dashed curve), and $T_3$(dotted curve).}
\end{center}
\end{figure}

In FIG.3, the entanglement degrees at three revival times $T_{1}$, $T_{2}$
and $T_{3}$ are compared as a function of parameter $\lambda _{0}$. 
Then we conclude that the entanglement degree takes a higher value as the subscript $k$ in $T_{k}$ increases, which makes us conjecture the following inequality for every $k$:
$$I_{{\cal E}_{T_{k}}^{*}\rho }\left( \rho _{T_{k}}^{A},\rho
_{T_{k}}^{F}\right) \leq I_{{\cal E}_{T_{k+1}}^{*}\rho }\left( \rho
_{T_{k+1}}^{A},\rho _{T_{k+1}}^{F}\right) .$$

\end{document}